\documentclass{gGAF2e}
\DeclareMathAlphabet\mathbfcal{OMS}{cmsy}{b}{n}

\usepackage{xcolor}

\renewcommand{\Omega}{{\varOmega}}

\begin{document}

\jvol{00} \jnum{00} \jyear{2018} %\jmonth{February}

%\markboth{Taylor \& Francis and I. T. Consultant}{\LaTeXe\ guide for authors}
\markboth{L.K.~Currie and S.M.~Tobias}{Geophysical and Astrophysical Fluid Dynamics}

%\articletype{GUIDE}

\title{Convection-driven kinematic dynamos with a self-consistent shear flow}

\author{L. K. Currie${\dag}$$^{\ast}$\thanks{$^\ast$Corresponding author. Email: L.K.Currie@exeter.ac.uk
\vspace{6pt}}  and S. M. Tobias${\ddag}$\\\vspace{6pt}  ${\dag}$Department of Physics and Astronomy, Stocker Road, University of Exeter,
EX4 4QL, UK\\ ${\ddag}$Department of Applied Mathematics, University of Leeds, LS2 9JT \\\vspace{6pt}\received{v4.4 released October 2012} }

\maketitle

\begin{abstract}
It is widely accepted that astrophysical magnetic fields are generated by dynamo action. In many cases these fields exhibit organisation on a scale larger than that of the underlying turbulent flow (e.g., the  eleven-year solar cycle). The mechanism for the generation of so-called large scale fields remains an open problem. In cases where the magnetic Reynolds number ($Rm$) is small, dynamo-generated fields are coherent but at (the astrophysically relevant) high $Rm$, the fields are overwhelmed by small scale fluctuating field. Recently \citet{tc:2013} have shown that an imposed large scale shear flow can suppress the small scale fluctuations and allow the large scale temporal behaviour to emerge. Shear is also believed to modify the electromotive force by introducing correlations between the flow and the field. However in previous models at high $Rm$ the shear is often artificially imposed or driven by an arbitrary body force. Here we consider a simple kinematic model of a convective dynamo in which shear is self consistently driven by the presence of a horizontal temperature gradient (resulting in a thermal wind) and a rotation vector that is oblique to gravity. By considering a $ 2.5$-dimensional system, we are able to reach high $Rm$ so that the dynamo approaches the asymptotic regime where the growth rate becomes approximately independent of $Rm$. We find the flows studied here to be excellent small-scale dynamos, but with very little systematic behaviour evident at large $Rm$. We attribute this to being unable to self-consistently generate flows with both large (net) helicity and strong shear in this setup.

\begin{keywords} Convection; dynamos; shear
\end{keywords}

\end{abstract}

\section{Introduction}\label{intro}
Astrophysical magnetic fields often exhibit a remarkable degree of order despite the high levels of turbulence. From planets to galaxy clusters --- via stars, disks and galaxies --- systematic magnetic fields are often observed \citep[see e.g.][]{Parker:1979, Moffatt:1978}, yet no self-consistent theory for their generation is currently agreed upon. It is generally accepted though that these magnetic fields are maintained by dynamo action, where inductive processes within the body overcome the ohmic diffusion \citep{tcb:11}. Moreover it seems to be possible for dynamo action to be efficient even at the astrophysically relevant parameter regime of high magnetic Reynolds number $Rm = U L / \eta$, where $U$ is a characteristic velocity $L$ a typical lengthscale and $\eta$ the magnetic diffusivity of the medium \citep{chilgil:1995}; typically $Rm = \mathrm{O}(10^8)$ in stars and even higher in galaxies.

A central issue is that, although dynamos can and do work at high $Rm$, all other things being equal their natural tendency is to generate magnetic field on a small (fluctuation) length-scale given by $l \sim Rm^{-1/2} L_u$ where $L_u$ is a characteristic lengthscale for a velocity eddy \citep{tobcatt:2008a}. The field produced tends to be random and small-scale rather than having systematic behaviour in either space or time like those observed in astrophysical objects. So how do systematic fields emerge? Arguments that go back to \citet{Parker:1955}, which were formalised by \citet{skr:1966} and \citet{kr:1980} as mean-field electrodynamics, demonstrate that systematic fields can only emerge owing to correlations between the flow and magnetic field that themselves emerge owing to breaking of reflectional symmetry in the system \citep[see also][]{Moffatt:1978}. This breaking of symmetry is often characterised in terms of the degree of kinetic helicity in the system. In general, because most astrophysical bodies are rotating and stratified, reflectional symmetry is naturally broken \citep{bransub:2005} and correlations and therefore systematic fields are expected to emerge.

In order to generate a tractable theory, mean field electrodynamics is completely formulated within a {\it kinematic framework}, where the velocity $\bm{u}$ is prescribed and solutions for the magnetic field $\bm{B}$ are exponentially growing or decaying.
The correlations discussed above emerge through a formally linear electromotive force (emf) given by $\mathbfcal{ E} = \langle \bm{u} \times \bm{b} \rangle$ where angle brackets represents a spatial, temporal or ensemble average \citep[e.g.,][]{dormysoward:2007} and $\bm{b}$ is the fluctuation of the magnetic field about that average. However, theoretical considerations and numerical calculations seem to suggest that at high $Rm$ the systematic magnetic field generated by the correlations in the emf may be swamped by the fluctuations that are inherent to turbulent dynamos \citep[e.g.][]{catthughes:06}. There has therefore been much effort to understand under what circumstances the systematic behaviour can win out over the fluctuations. Within the kinematic framework (with which this paper is concerned --- we shall return to nonlinear considerations in the conclusions) it has been suggested that the presence of a large-scale shear flow (a natural flow in many astrophysical fluids) could remedy the natural tendency of the fluctuation dynamo to triumph over the systematic dynamo \citep{tc:2013}. There are two mechanisms that have been proposed by which this may be achieved. The first is that the shear flow enhances the correlations between the flow and the field leading to the production of a net emf, thus helping the large-scale dynamo. There are many papers \citep{Yousef08,Kapya09,Sridhar10} that demonstrate this effect for imposed shear flows and turbulence at low $Rm$ and attempt to characterise the nature of the new correlations by relating them back to the large-scale field via transport coefficients (this is possible in a purely kinematic/linear framework). The second route by which the systematic dynamo may win out is for the shear to remove the turbulent fluctuations of the magnetic field without having a significant effect on the correlations that lead to the systematic dynamo. In a series of papers \citep{tc:2013,tc:2015,npct:2017} it has been shown that, for a range of model ($ 2.5$-dimensional) flows at high $Rm$, the primary effect of a shear flow is to reduce the growth rate of the fluctuation dynamo (leaving the mean emf largely unaffected). For flows with enough breaking of reflectional symmetry (characterised by net helicity in the flow) this can lead to the unambiguous observation of systematic dynamo action (indeed oscillatory dynamo action) at high $Rm$.

In the papers described above, the role of shear was investigated for shear flows imposed with a prescribed strength. In astrophysical situations the shear flows usually self-consistently  emerge via the interaction of turbulence with rotation leading to correlations in the flow \citep[see e.g.][and the references therein]{brunbrowning:2017}. This situation is much less widely studied at high $Rm$, though of course there are a large number of spherical convection dynamos that are investigated at lower $Rm$ where differential rotation and magnetic fields emerge self-consistently \citep{pc:2014,abmt:2015}. In this paper we consider a simple model of the interaction of convection with rotation leading to shear flows and dynamo action in order to characterise under what circumstances (if any) systematic dynamo action emerges. Our model is similar in spirit, though different from, that considered by \citet{PontGilSow:2001}, which described the interaction of convection and dynamos with an Ekman spiral flow.

\section{Model setup}\label{setup}
We consider a plane layer of height $d$ of Boussinesq fluid rotating about an axis that is oblique to gravity. We assume that gravity $\bm{g}=(0,0,-g)$ is constant and acts in the negative $z$ (downwards) direction, with $x$ pointing eastwards and $y$ northwards. 
The rotation vector lies in the $y$--$z$ plane and is given by $\bm\Omega=\Omega(0,\cos\phi,\sin\phi)$, where $\Omega$ is the rotation rate and $\phi$ is the latitude. Throughout this paper, we will consider $0<\phi\leq \pi/2$ so that this local model is representative of a region at latitude $\phi$ in the northern hemisphere of a spherical body hence when $\phi=\pi/2$, the rotation is vertical.
Convection is driven through an adverse vertical temperature gradient and in addition, an imposed temperature gradient in the $y$ direction produces a thermal wind shear \citep{htg:1980}. In this case, the dimensionless temperature basic state is given by
\begin{equation}\label{Tbasic}
T_0\,=\,T_c+T_yy-z\,,
\end{equation}
where $(\upartial T/\upartial z)d$ has been used as the characteristic unit of temperature and $d$ the characteristic unit of length. $T_c$ is a constant and $T_y$ is a dimensionless measure of the horizontal temperature gradient. For $T_y<0$, the temperature increases southwards (equatorwards), as is the case on Earth, for example. In this paper, we restrict our attention to this case and do not consider what happens when $T_y> 0$.
A poleward pressure gradient is produced by the horizontal temperature gradient and is balanced by the Coriolis force acting on a thermal wind, this leads to a velocity basic state with vertical shear (the thermal wind) and is given by $\bm{U_0}=(U_0(z),0,0)$, where \citep{htg:1980}
\begin{equation}\label{Ubasic}
U_{0}\,=\,-\,\frac{T_yRa}{Ta^{{1}/{2}}\sin\phi}\left(z-\tfrac{1}{2}\right),
\end{equation}
for $\phi\neq0$.
This expression involves the following standard dimensionless numbers: the Rayleigh number $Ra=\alpha g d^4 \upartial T / \upartial z / \kappa\nu$ and the Taylor number $Ta=4\Omega^2d^4 / \nu^2$, where $\alpha$ is the coefficient of thermal expansion, and $\kappa$ and $\nu$ are the thermal and viscous diffusivities respectively.
Such a hydrodynamic system can drive a flow $\bm{u}$ which may or may not be capable of generating a magnetic field through dynamo action.
We determine if any magnetic field $\bm{B}=(B_x,B_y,B_z)$ is generated from the resulting flows by solving the induction equation. Here, we only consider the kinematic dynamo problem and do not account for any back reaction of the field on the flow. 
We consider velocity perturbations $\bm{u}$ and temperature perturbations $\theta$ to the basic state given by (\ref{Tbasic}) and (\ref{Ubasic}). In this case, the dimensionless governing equations are given by \citep[see e.g.,][]{curthesis:2014}
\begin{gather}
%\begin{equation}\label{incomp}
\bm\nabla{\bm \cdot}\bm{u}\,=\,0\,,\label{incomp}\\
%\end{equation}
%\begin{equation}
\frac{\upartial\bm{u}}{\upartial t}+U_0\frac{\upartial \bm u}{\upartial x}+w\frac{{\rm d}\bm{U_{0}}}{{\rm d} z} -Pr\nabla^2\bm{u}+Ta^{1/2}Pr\bm\Omega\times\bm{u} +Pr\bm\nabla p -RaPr\theta\bm{\hat e_z}\hskip 30mm\nonumber\\
\hskip 50mm  =\,-\,(\bm{u}{\bm \cdot}\bm\nabla)\bm{u}\,,\\
%\end{equation}
%\begin{equation}\label{heateq}
\frac{\upartial\theta}{\upartial t}+ U_0\frac{\upartial\theta}{\upartial x}-\nabla^2\theta-w+T_yv\,=\,-\,(\bm{u}{\bm \cdot}\bm\nabla)\theta\,,\label{heateq}\\
%\end{equation}
%\begin{equation}
\bm\nabla{\bm \cdot}\bm{B}\,=\,0\,,\\
%\end{equation}
%\begin{equation}\label{indeq}
\hskip 5mm \frac{\upartial \bm{B}}{\upartial t}-\zeta\nabla^2\bm{B}-\bm\nabla \times(\bm{U_{0}}\times\bm{B})\,=\,\bm\nabla \times(\bm{u}\times\bm{B})\,,\label{indeq}
%\end{equation}
\end{gather}
where $p$ denotes the pressure perturbation. We have used the thermal diffusion time $d^2 / \kappa$ as the characteristic unit of time; velocities therefore have a typical scale given by $\kappa/d$. $Pr=\nu / \kappa$ is the Prandtl number (ratio of viscous to thermal diffusivities) and $\zeta=\eta / \kappa$ is the ratio of magnetic to thermal diffusivities. The more conventional magnetic Prandtl number $Pm$ can be obtained from $Pm=Pr / \zeta$.

For this study, we consider so-called $2.5$-dimensional flows so that the velocity field contains all three components but each component depends only on the two spatial components, $y$ and $z$, i.e., $\bm{u}=(u(y,z),v(y,z),w(y,z))$. It follows that $\theta$ and $p$ are also only functions of $y$ and $z$ and so all $x$-derivatives in equations (\ref{incomp})-(\ref{heateq}) can be taken to be zero. The magnetic field, however, must depend on all three spatial coordinates in order to avoid anti-dynamo theorems \citep{cowling:1933}. By considering simplified hydrodynamic flows we are able to reach much higher $Rm$ and probe dynamo action in a more astrophysically relevant regime.

We assume the top and bottom boundaries to be: held at fixed temperature, impermeable, perfectly conducting and at fixed stress (so that the perturbations to the thermal wind are stress free), that is
\begin{equation}
\theta=\frac{\upartial u}{\upartial z}\,=\,\frac{\upartial v}{\upartial z}\,=\,w\,=\,\frac{\upartial B_x}{\upartial z}\,=\,\frac{\upartial B_y}{\upartial z}\,=\,B_z\,=\,0 \hskip 10mm \text{on $z=0,1$.}
\end{equation}
Furthermore, we assume all quantities to be periodic in the $y$ direction. Note, although the basic state temperature (\ref{Tbasic}) does not satisfy periodic boundaries in $y$, only the gradient of the basic state temperature appears in the governing equations.

Our assumption that $\bm{u}$ depends only on $y$ and $z$ (and $t$) means that the induction equation (\ref{indeq}) is separable and solutions can be written as
\begin{equation}
\bm{B}\,=\,\bm{\hat B}(y,z,t)\exp\bigl({\rm i}k_x x\bigr)\,,
\end{equation}
where $\bm{\hat B}=(\hat B_x, \hat B_y, \hat B_z)$ is a complex amplitude and $k_x$ is the wavenumber in $x$. With this, the system of equations (\ref{incomp})-(\ref{indeq}) becomes two-dimensional allowing larger $Rm$ to be reached with relative computational ease.
We solve the governing equations using the open-source, pseudospectral code Dedalus (K. J. Burns et al. 2018, in preparation).
A domain size of $[L_y,L_z]=[10, 1]$ is used throughout, with resolutions of up to $4096 \times 512$ modes (after dealiasing) utilised. The large horizontal box size ensures a separation of scales between the smaller-scale turbulence and the box size.

\subsection{Diagnostic quantities}\label{diagquant}
To assess the hydrodynamic properties of the flows satisfying (\ref{incomp})-(\ref{heateq}), we define the following quantities.
The relative helicity is given by
\begin{equation}\label{relhel}
h(z)\,=\left\langle\frac{\langle \bm u' {\bm \cdot} \bm \omega' \rangle_{y}}{\langle \bm u'^2\rangle_y^{{1}/{2}}\langle \bm \omega'^2\rangle_y^{{1}/{2}}}\right\rangle_{\!\!t}\,,
\end{equation}
where $\bm \omega'=\bm\nabla\times\bm u'$ is the vorticity, and $\langle{\cdot}\rangle_y$, $\langle{\cdot}\rangle_t$ denote an average over $y$ and $t$ respectively. $\bm u'=\bm u-\langle\bm u\rangle_y$ is the fluctuation of $\bm u$ about its mean state $\langle\bm u\rangle_y$. We use $\bm u'$ in the calculation of $h$ because we are interested in the helicity of the turbulent eddies and not the large-scale component of the flow.
Since we assume the Boussinesq approximation, the system possesses a symmetry about the midplane ($z=0.5$) and therefore an average of $h$ over $z$ would lead to zero net helicity. So instead, we define $H=\langle h(z) \rangle_{z_{-}}$ where the average in $z$ is taken over the lower half plane only ($z\leq0.5$). For a statistically steady system, we would expect this value to equal the average of $h(z)$ taken over the upper half plane. We shall return to the importance of helicity in the conclusions.

We consider two measures of the relative shear in the flow: firstly, we define
\begin{equation}\label{relshearu}
S_u\,=\left\langle \frac{KE_{utot}}{KE_{tot}} \right\rangle_{\!t}\,,
\end{equation}
where $KE_{utot}=0.5\langle (\langle (u+U_0) \rangle_y)^2 \rangle_z $ is the kinetic energy in the $x$ component of the total mean flow and $KE_{tot}=0.5 \langle (u+U_0)^2+v^2+w^2) \rangle_{yz}$ is the total kinetic energy in the fluctuations and the basic state velocity. 
Secondly, we define
\begin{equation}\label{relshear}
S\,=\left\langle \frac{KE_{utot} + KE_v}{KE_{tot}} \right\rangle_{\!t}\,,
\end{equation}
where $KE_{v}=0.5\langle (\langle v \rangle_y)^2 \rangle_z $ is the kinetic energy in the $y$ component of the mean flow. These quantities give a measure of the energy in the shear flow relative to the total kinetic energy. 
\\

\section{Hydrodynamic flows}\label{hydro}
\begin{figure}
\begin{center}
\includegraphics[trim={0mm 0mm 0mm 0mm}, clip, scale=0.96]{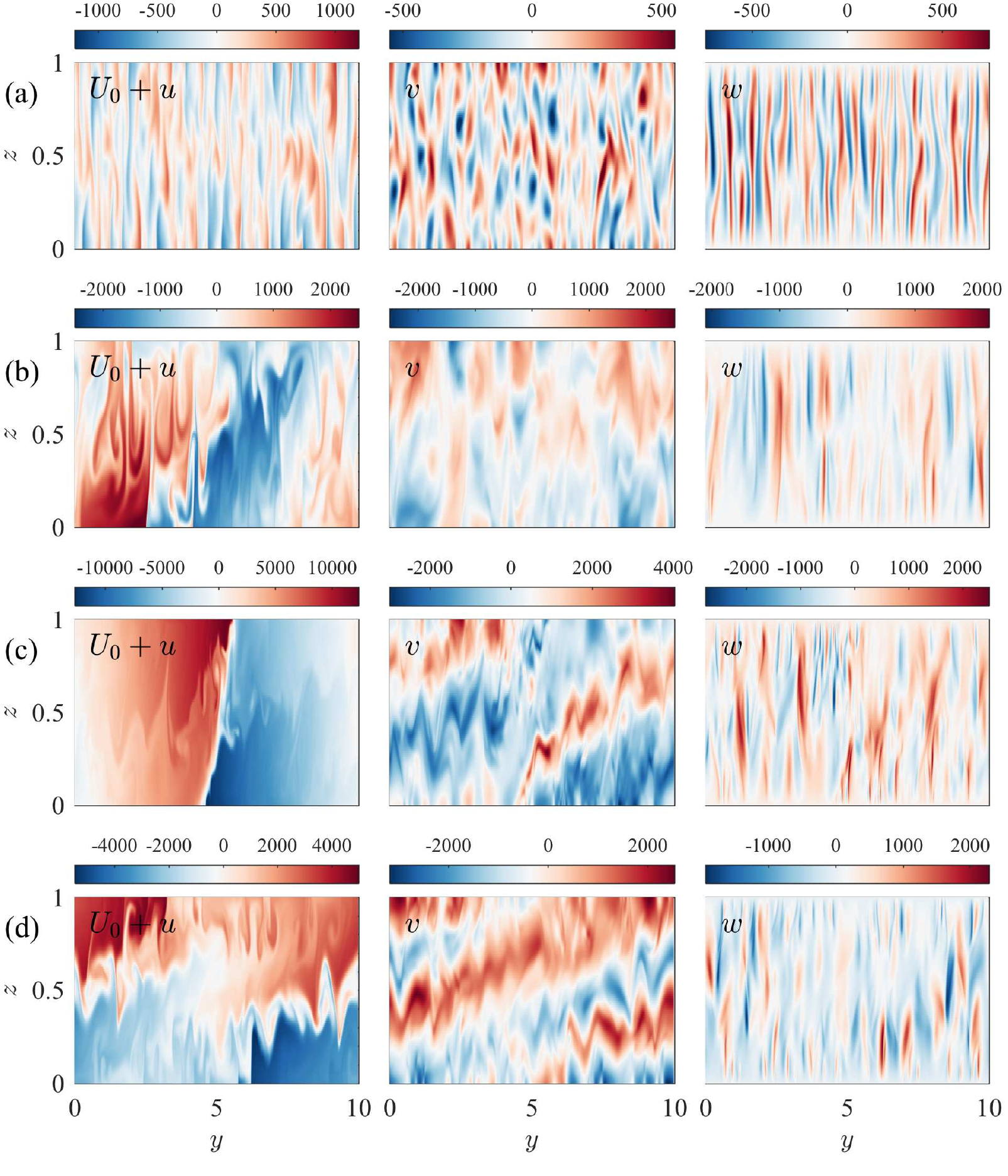}
\caption{\label{fig1} %
Snapshots of $u_{total}=U_0+u$ (first column), $v$ (second column) and $w$ (third column) for cases with $N_{crit}=11.63$, $Ta=5\times10^6$ and (a) $T_y=0$, $\phi=\pi /2$, (b) $T_y=0$, $\phi=\pi /12$, (c) $T_y=-5$, $\phi=\pi /2$ and (d) $T_y=-5$, $\phi=\pi /12$. (Colour online)}%
\end{center}
\end{figure}

Flows governed by equations (\ref{incomp})-(\ref{heateq}) are determined by the dimensionless parameters $Pr$, $Ta$, $\phi$, $Ra$ and $T_y$. In this paper we will consider, for simplicity, only fluid with $Pr=1$. $Ta$ will also be fixed, at $5\times10^6$ (to see the effect of rotation rate on the flows see e.g., \citealt{htg:1980,curthesis:2014}). We then vary the latitude, Rayleigh number and thermal wind strength to achieve a variety of flows. The degree of supercriticality of the flows is measured through $N_{crit}=Ra/Ra_c$, where $Ra_c$ is the value of $Ra$ required for convection to onset in a finite box of size $[L_y, L_z] = [10, 1]$. The effect of $T_y$ on $Ra_c$ is investigated in \citet{curthesis:2014}, but in all cases considered here a larger $|T_y|$ corresponds to a smaller $Ra_c$.
The difference in flow morphology in four different regimes (at the pole, without and with thermal wind, and near the equator, without and with thermal wind (but all with $N_{crit}=11.63$)) is shown by the snapshots of the velocity field in each of these regimes (see figure~\ref{fig1}). In (a), $T_y=0$ and the rotation is vertically aligned (i.e., $\phi=\pi/2$), in (b) $T_y=0$ but now the layer is close to the equator with $\phi=\pi /12$ and so the convection rolls are tilted to align with the rotation vector. A key difference between (a) and (b) can perhaps be attributed to the reduction in rotational constraint that comes from decreasing $\phi$ since this decreases the magnitude of the vertical component of rotation; the horizontal length scale of the solution is larger in (b) and the flow velocities are higher. The addition of a strong thermal wind introduces a strong shear in the layer, which is particularly evident in the zonal ($x$-component) velocity of cases (c) and (d). The vertical alignment of the shear appears to depend on $\phi$.

\begin{figure}
\begin{center}
\includegraphics[trim={0mm 0mm 0mm 0mm}, clip, scale=0.98]{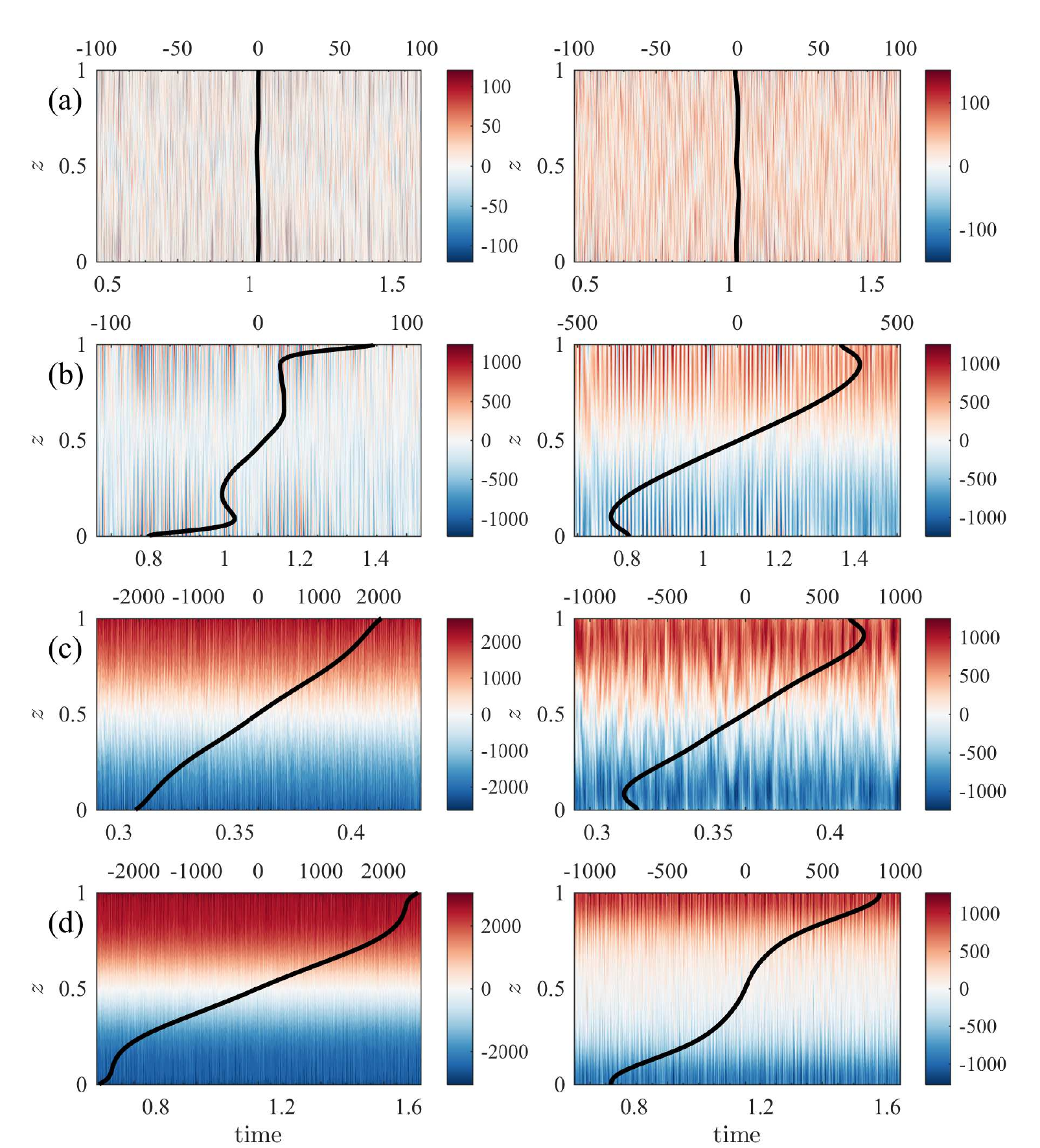}
\caption{\label{fig2} %
Mean flow components $\langle u_{total}\rangle_y=\langle U_0+u \rangle_y$ (left) and $\langle v \rangle_y$ (right) for the same cases as in figure~\ref{fig2}. The contours give the mean flow as a function of depth and time (bottom axes and colour bar) and the thick solid line gives the time averaged mean (top axes). (Colour online)}%
\end{center}
\end{figure}

The corresponding mean ($y$-averaged) flows are shown in figure~\ref{fig2}. The contour plots give the mean flows as a function of $z$ and time while the overlying black line gives the time-averaged values of the mean flow. Clearly in case (a), whilst there is a non-zero mean flow at each instance in time, on averaging over a long enough period to achieve steady statistics, the mean flows are very small (this is to be expected as there is no preferred direction in the horizontal plane). In (b), the tilting of the rotation now breaks this symmetry and this leads to systematic mean flows \citep[such flows have been seen in many cases, e.g.,][]{hatsom:1983, bht:1998, julienknobloch:1998, curtob:2016}. The addition of a basic state thermal wind leads to a significant increase in the strength of the mean flows, this is most obvious in the zonal direction as, by definition, the mean flow in this direction contains the basic state flow, but from (c) and (d) right-hand panels, it is also clear that the addition of a thermal wind leads to strong mean flow in the $y$ direction also.

\begin{figure}
\begin{center}
\includegraphics[trim={0mm 0mm 0mm 0mm}, clip, scale=0.98]{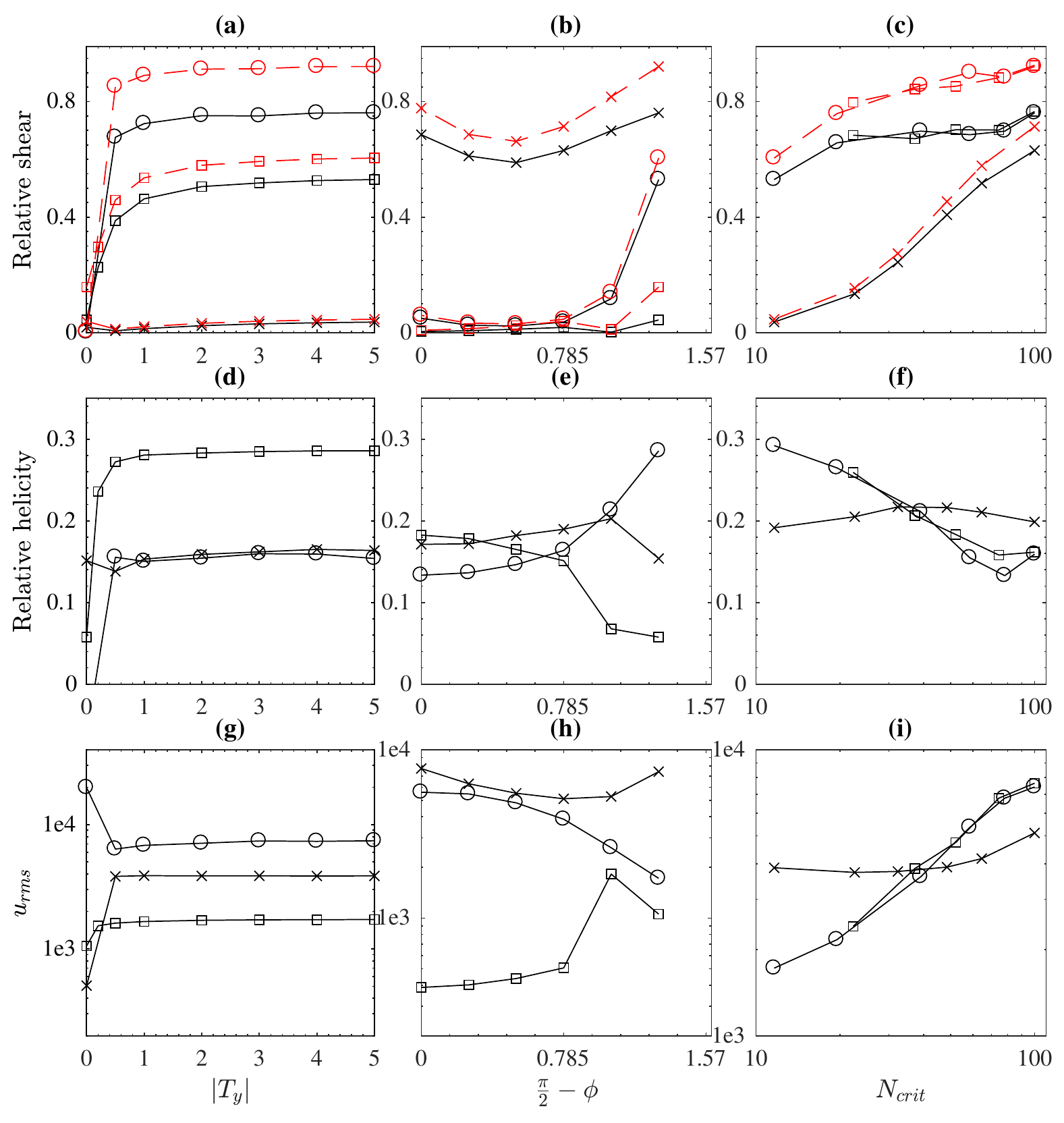}
\caption{\label{fig3}
\emph{Top row}: Measures of the relative shear $S_u$ (red, dashed) and $S$ (black, solid) as defined by (\ref{relshearu}) and (\ref{relshear}) respectively. \emph{Middle row}: Relative helicity, $H$, as defined in section \ref{diagquant} and \emph{bottom row}: rms measure of the velocity perturbations, $u_{rms}$. Different marker types indicate different parameter regimes. In the first column $|T_y|$ is varied for fixed $\phi$ and $N_{crit}$; squares represent $\phi=\pi /12$, $N_{crit}=11.63$, circles represent $\phi=\pi / 12$, $N_{crit}=100$ and crosses represent $\phi=\pi /4$, $N_{crit}=11.63$. In the second column $\phi$ is changed for fixed $T_y$ and $N_{crit}$. Note, on the $x$-axis we plot $\pi/2-\phi$ so that moving along the $x$-axis corresponds to tilting the rotation vector further from the vertical (decreasing latitude, $\phi$); squares represent $T_y=0$, $N_{crit}=11.63$, circles represent $T_y=-5$, $N_{crit}=11.63$ and crosses represent $T_y=-5$, $N_{crit}=100$. In the third column, $N_{crit}$ is changed whilst keeping $T_y$ and $\phi$ constant; squares represent $T_y=-200$, $\phi=\pi/12$, circles represent $T_y=-5$, $\phi=\pi/12$ and crosses represent $T_y=-5$, $\phi=\pi/4$.}
\end{center}
\end{figure}

As was discussed in section \ref{intro}, shear and helicity are expected to play an important role in magnetic field generation through dynamo action, we therefore consider the size of the relative shear and helicity for different types of flow (some examples of which were sampled in figures \ref{fig1} and \ref{fig2}). 
The relative shear, $S$ (defined in (\ref{relshear})) is shown in figure~\ref{fig3}(a)-(c); $S_u$ (defined in (\ref{relshearu})) is also shown but the trends are found to be very similar and so we restrict ourselves to describing the behaviour of $S$ in the text. From (b) we see that, although $S$ increases at lower latitudes (note, the latitude decreases with increasing $x$-axis) it is relatively small when $T_y=0$  -- taking values of less than approximately 16\%. Introducing a thermal wind can increase $S$ significantly, particularly at lower latitudes (though the exact behaviour depends crucially on the other system parameters, e.g., $Ra$). Once a moderate thermal wind is present, increasing $T_y$ further only acts to increase $S$ slightly; in (a), $S$ increases with $|T_y|$ until approximately $T_y=-1$ where $S$ then remains roughly constant as $T_y$ is decreased further from $-1$ to $-5$. Furthermore, (c) shows that even with $T_y=-200$, $S$ does not change much from the value at $T_y=-5$ (all other parameters being equal). We note that $T_y=-200$ is an extremely large horizontal temperature gradient, it is unphysically 200 times that of the vertical temperature gradient imposed to drive convection; we include it here merely for demonstrative purposes.
Figure~\ref{fig3}(b) shows that tilting the rotation further from the vertical can lead to a non monotonic behaviour when a thermal wind is present; as the rotation vector is tilted from the vertical, $S$ decreases before increasing at low latitudes (the case when $T_y=0$ is less clear). In both thermal wind cases considered in (b) (crosses and circles) $S$ is largest near the equator.
Figure~\ref{fig3}(c) highlights that as $N_{crit}$ is increased, $S$ increases; this is because both the mean flow driven by the turbulence and the basic state thermal wind are stronger.

Figure~\ref{fig3}(d)-(f) depict how the relative helicity, $H$ (defined in section~\ref{diagquant}) changes as a function of $T_y$, $\phi$ and $N_{crit}$. Clearly the behaviour is complex and depends strongly on which area of parameter space one is examining. However, in all cases considered here, $H$ never appears to exceed 0.3, (recall this a half-box average and that the net helicity is close to zero). In (d), similar to the behaviour of $S$, $H$ does not change much with $T_y$ for $T_y<-1$ (all other parameters fixed).
For $T_y=0$ and $N_{crit}=11.63$ (see (e), squares) $H$ decreases as the tilt angle increases; this is likely to be because of the reduction in rotational constraint, allowing for larger velocities (see figure~\ref{fig3}(h)).
When $T_y\neq0$, the situation is more complicated and depends on $N_{crit}$. For example, when $N_{crit}=11.63$ (circles), $H$ increases as $\phi$ decreases but for $N_{crit}=100$ (crosses), $H$ is not monotonic and remains approximately constant. In both cases, the behaviour of $H$ is well correlated with the root-mean-square perturbation velocity, $u_{rms}$: an increase in $H$ coincides with a decrease in $u_{rms}$ and vice versa.
Figure~\ref{fig3}(f) highlights that $H$ can both increase and decrease with $N_{crit}$ depending on the value of $\phi$, but for $T_y\neq0$, the value of $H$ is largely independent of $T_y$ itself.
The corresponding values of $u_{rms}$ are given in figure~\ref{fig3}(g)-(i). In (g) and (h), $N_{crit}$ is constant for each set of parameters, however, the value of $u_{rms}$ changes with $\phi$ and whether a thermal wind is present or not.

\begin{figure}
\begin{center}
\includegraphics[trim={0mm 0mm 0mm 0mm}, clip, scale=0.96]{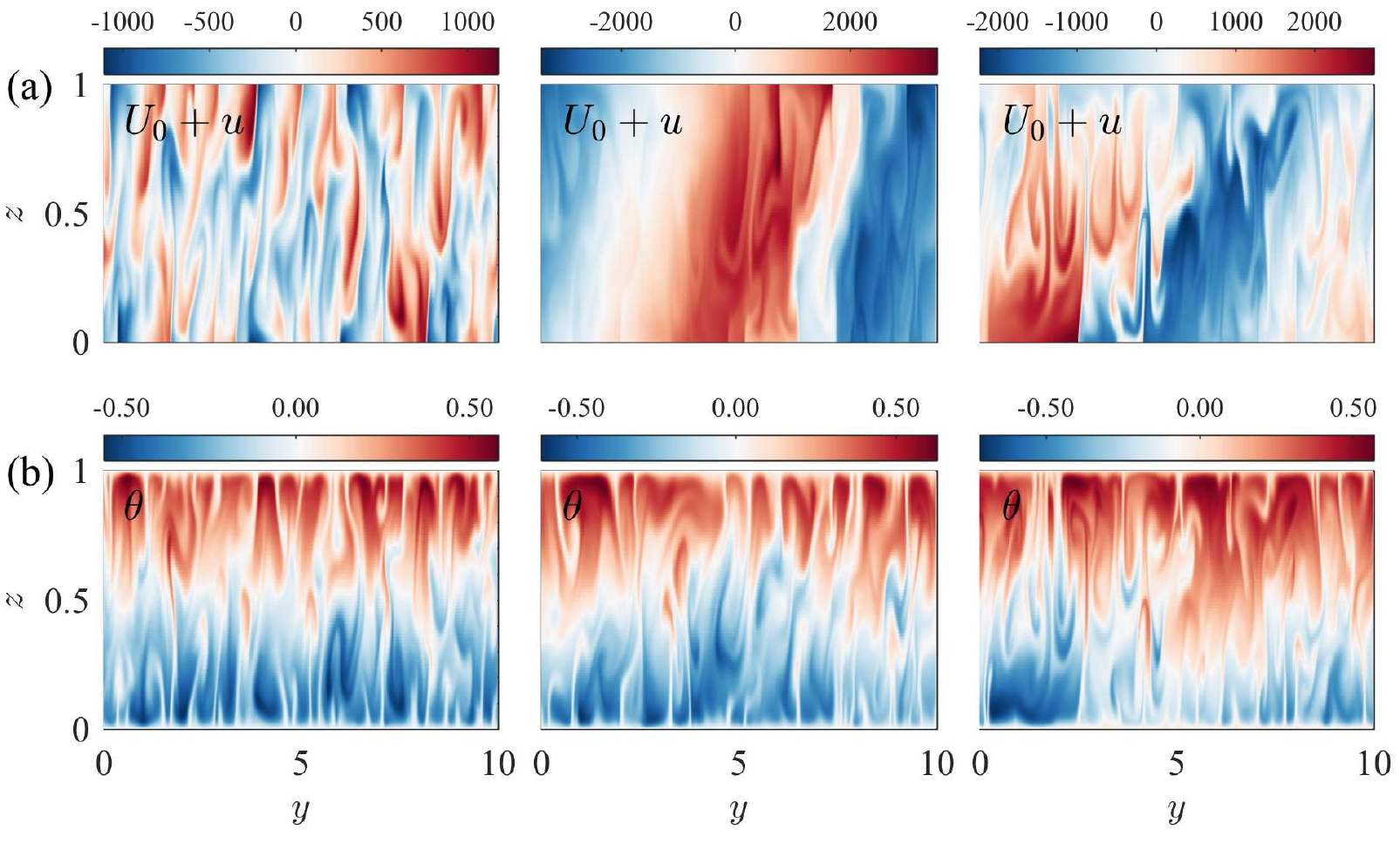}
\caption{\label{fig4}
Snapshots of (a) $u$ and (b) $\theta$ for $T_y=0$ and $N_{crit}=11.63$ for cases with $\phi=\pi/4$, $\phi=\pi/6$ and $\phi=\pi/12$ from left to right respectively. (Colour online)}
\end{center}
\end{figure}

We finish describing this figure by commenting on the distinctive behaviour of the case when $\phi=\pi/6$, $T_y=0$ and $N_{crit}=11.63$; in this case, $u_{rms}$ appears to be significantly out-of-trend when comparing with the same $T_y$, $N_{crit}$ but varying $\phi$ (see figure~\ref{fig3}(h), squares). To understand this further, we examine the spatial form of the velocity and temperature perturbations in this case (see figure~\ref{fig4}). Whilst the temperature perturbation $\theta$ is very similar in all three cases shown, $u$ shows significant differences in its spatial structure.
As $\phi$ is decreased from $\phi=\pi/4$ to $\phi=\pi/6$, the length scale of $u$ increases significantly and $u$ itself is less turbulent for $\phi=\pi/6$ than for $\phi=\pi/4$. We suggest that this is a trait of these parameters that allows a more laminar solution to be dominant at $\phi=\pi/4$, with this particular solution having a higher $u_{rms}$, as shown in figure~\ref{fig3}(h). Decreasing $\phi$ further to $\phi=\pi/12$ leads to a more turbulent solution again (although not as turbulent as when $\phi=\pi/4$) and this coincides with $\phi=\pi/12$ having an $u_{rms}$ that is less than that when $\phi=\pi/6$ but bigger than that when $\phi=\pi/4$.

The variety of flows studied here have different heat transport properties, a more detailed description of the heat transport across the layer can be found in Appendix \ref{app1}.

\section{Kinematic dynamo action}\label{dynamo}
\subsection{Growth rate}
\begin{figure}
\begin{center}
\includegraphics[trim={0mm 0mm 0mm 0mm}, clip, scale=0.96]{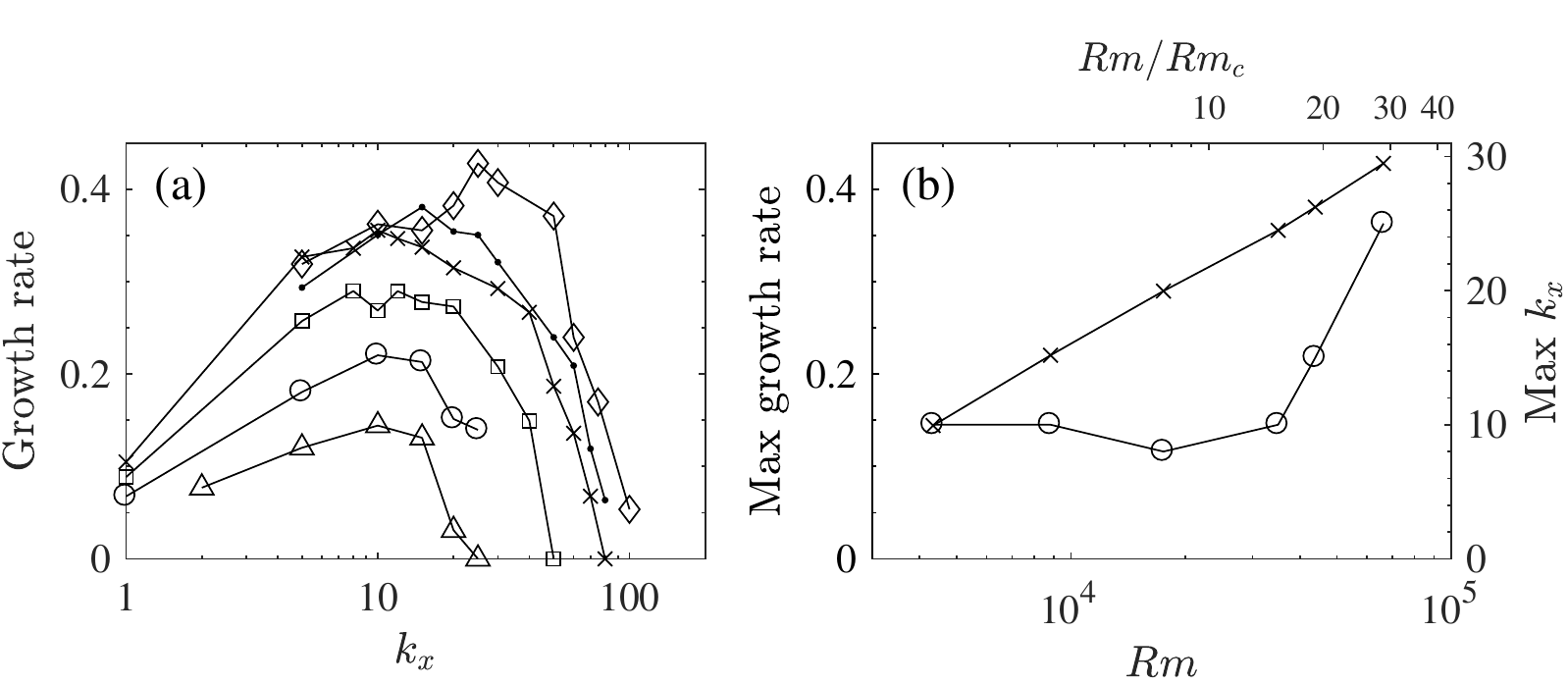}
\caption{\label{fig5}(a) Growth rate against $k_x$ for $Rm\sim4300$ (triangles), $Rm\sim8700$ (circles), $Rm\sim17000$ (squares), $Rm\sim35000$ (crosses) $Rm\sim44000$ (dots) and $Rm\sim66000$ (diamonds), for cases with $\phi=\pi /12$, $T_y=-5$ and $N_{crit}=19.53$. Negative growth rates are plotted with a zero growth rate. (b) Maximum growth rate (crosses) and corresponding $k_x$ (circles) against $Rm$ for the same $\phi$, $T_y$ and $N_{crit}$ as in (a). The growth rates have been scaled to be be in units of the turnover time of the flow.}
\end{center}
\end{figure}

The turbulent flows that we examined in section \ref{hydro} are expected to be very good dynamos even at high $Rm$ as they are very time-dependent and  chaotic \citep{Vishik:1989,Klapper:1995,chilgil:1995}. As discussed in section~\ref{hydro} they also have self-consistently generated shear flows, which may play an important role in modifying the dynamo properties. In this section, we examine the properties of magnetic field, such as spatial dependence and growth rate, that is generated via kinematic dynamo action. As described in section \ref{setup}, the induction equation is solved with the horizontal wavenumber $k_x$ as a parameter and, because of the linearity of the induction equation, the magnetic field will either grow exponentially (if it is a dynamo) or it will decay exponentially (on average). In any case, a well-defined average growth rate can be determined by calculating a line of best fit for the exponentially growing magnetic energy in the time-series.  Each $k_x$ will have an associated growth rate and we are interested in which mode gives the maximal growth rate \citep[see, for example][]{Roberts:1972,Galloway:1992}. Figure~\ref{fig5}(a) shows the growth rate as a function of $k_x$ for six different values of $\zeta$ (and therefore $Rm$, since here $Rm=u_{rms} / \zeta$). Each curve has a well defined preferred $k_x$ and for high $k_x$ the growth rate becomes negative.
In figure~\ref{fig5}(b) we plot the maximum growth rate and corresponding $k_x$ as a function of $Rm$. The fact that the growth rate continues to increase with $Rm$ indicates we are not in an asymptotic regime where the dynamo reaches an asymptotic $\mathrm{O}(1)$ growth rate (as measured in units of the turnover time of the flow)  as $Rm \rightarrow \infty$ \citep{chilgil:1995}; hence it is not possible to say whether the dynamos here are fast. A \textit{quick} dynamo reaches its asymptotic growth rate close to $Rm_c$ where $Rm_c$ is the critical $Rm$ \citep{tobcatt:2008a}. That is, it approaches asymptoticity for $\chi = Rm/Rm_c$ not too large (maybe $\mathrm{O}(10)$) \citep{tc:2015}. In figure~\ref{fig5}(b) the largest $Rm$ have $\chi\sim30$ and as discussed above we are not in the asymptotic regime, therefore the dynamos in this case appear not to be quick either.

\begin{figure}
\begin{center}
\includegraphics[trim={0mm 0mm 0mm 0mm}, clip, scale=0.98]{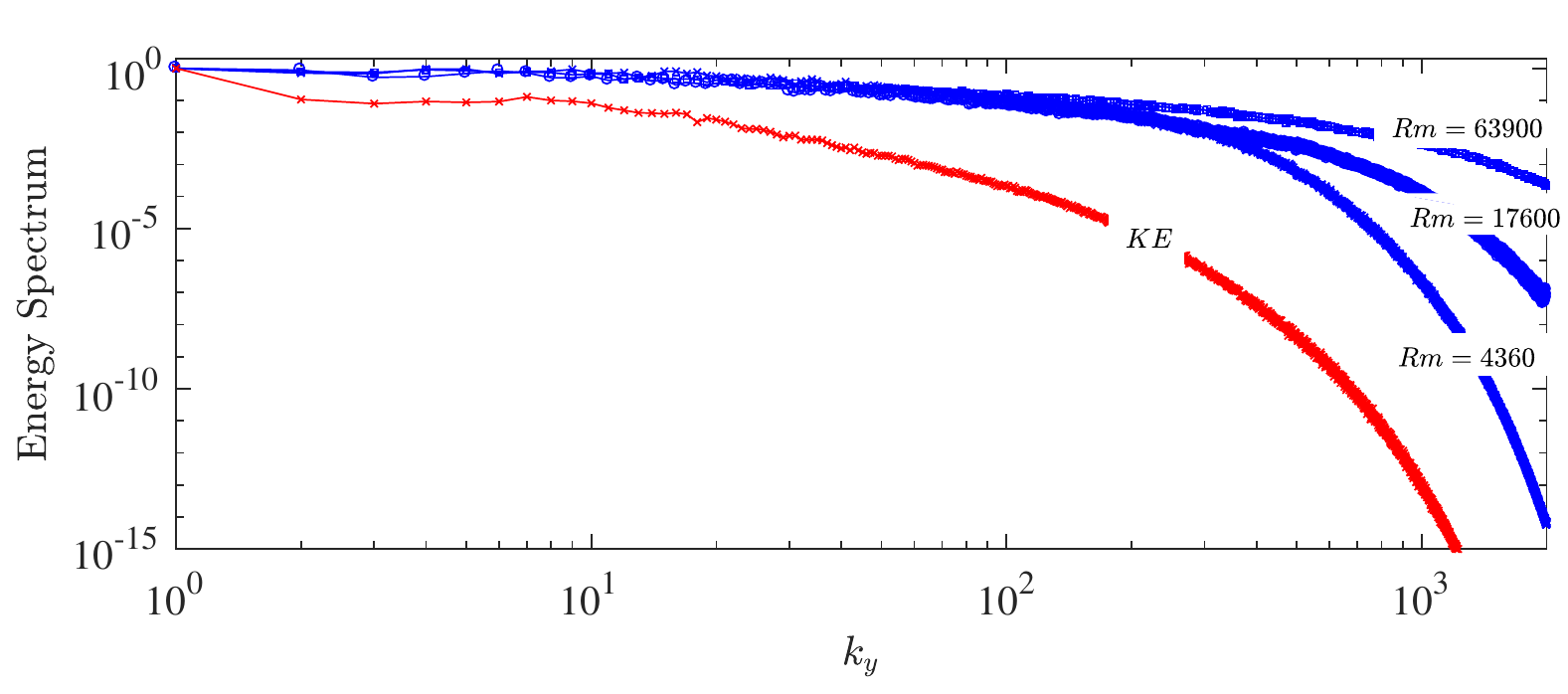}
\caption{\label{fig6} Spectrum of the magnetic energy (blue) for three different $Pm$ (equivalently $Rm$). The kinetic energy spectrum is the same in all three cases and is given by the red line. All spectra have been normalised to be 1 at their maximum value. (Colour online)}
\end{center}
\end{figure}

We note that as we decrease $\zeta$ to increase $Rm$ (holding the flow fixed) we are increasing the magnetic Prandtl number $Pm$. All of these dynamos are in the $Pm > 1$ regime with some of these having $Pm \gg 1$ (in figure~\ref{fig6}, $Pm$ ranges from 2 to 30). It is well-known that for such dynamos the dissipative cut-off for the magnetic field lies to the right of that for the velocity in $k$-space (as demonstrated by the spectra in figure~\ref{fig6}). Here we can see the dissipative cut-off moving progressively to the right as $Pm$ is increased. We note here that in many convectively driven dynamos (for example the dynamos of planetary and stellar interiors) the correct parameter regime has $Pm \ll 1$, with the dissipative cut-off for the magnetic field to the left of that for the velocity. This regime is hard to access numerically \citep[see e.g.][]{2007NJPh....9..300S} for dynamo action, though magnetoconvection calculations in this parameter regime may prove of interest.


\begin{figure}
\begin{center}
\includegraphics[trim={0mm 0mm 0mm 0mm}, clip, scale=1]{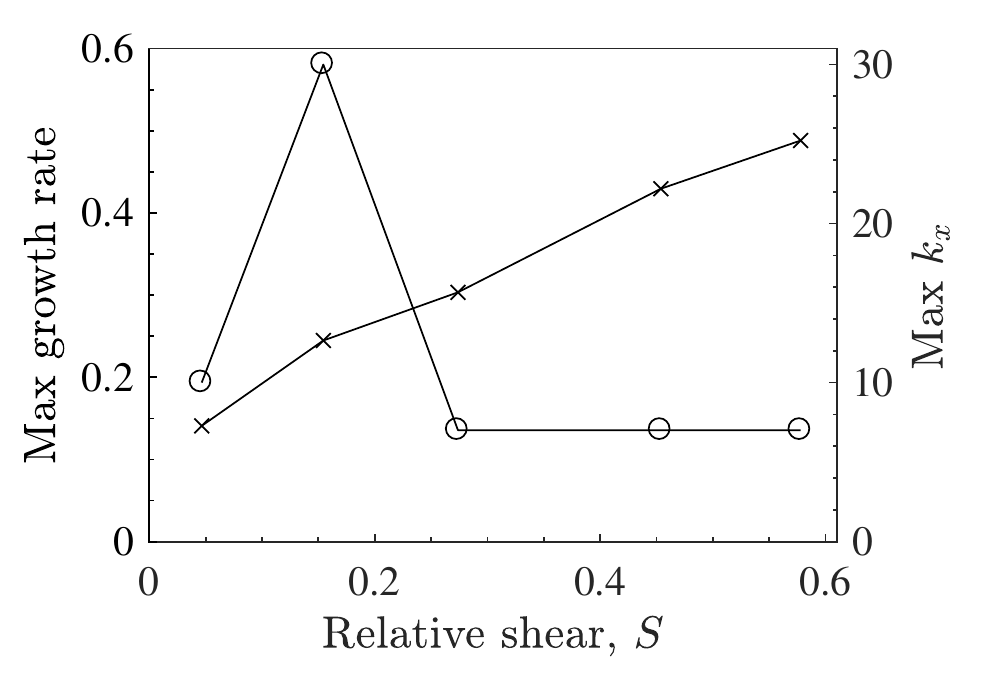}
\caption{\label{fig7} Maximum growth rate (crosses) and corresponding $k_x$ (circles) as a function of the relative shear, $S$, for $T_y=-5$ and $\phi=\pi/4$. $S$ is increased by increasing $N_{crit}$ between 11.6 and 64.6 -- see figure \ref{fig3}(c). In all cases $\zeta=1/16$.}
\end{center}
\end{figure}

To determine how the growth rate of the dynamo is affected by the shear in the flows, we consider how the maximum growth rate changes with $S$ (defined in (\ref{relshear})) -- see figure \ref{fig7}. For this case, $Ta=5\times10^6$, $\phi=\pi /4$, $T_y=-5$ and $S$ is changed by varying $N_{crit}$ between 11.6 and 64.6 (the hydrodynamic properties of these flows were considered in figure~\ref{fig3}(c), (f) and (i)). From figure~\ref{fig3}(i), we see that $u_{rms}$ remains roughly constant and so for fixed $\zeta=1/16$ (as is the case in figure~\ref{fig7}) $Rm$ is also roughly constant. However, $Rm_c$ may change as the $N_{crit}$ is varied. Clearly, the maximum growth rate increases with $S$, this is the expected behaviour at low $Rm$ where shear is thought to aid small-scale dynamo action \citep{Yousef08}; it is also another indicator that we are not in the asymptotic regime where it is expected the shear would decrease the growth rate of the small-scale dynamo \citep{tc:2013}.
The wavenumbers corresponding to the maximum growth rates are smallest for the largest shears which agrees with the idea that strong shear will wipe out high $k$ (small-scale modes). However, as noted previously, since here the shear is self-consistently generated by the flow, to obtain different values of $S$ the flow has to vary and in general $Rm_c$ will vary too.

\subsection{Magnetic field}
\begin{figure}
\begin{center}
\includegraphics[trim={1mm 0mm 0mm 0mm}, clip, scale=0.97]{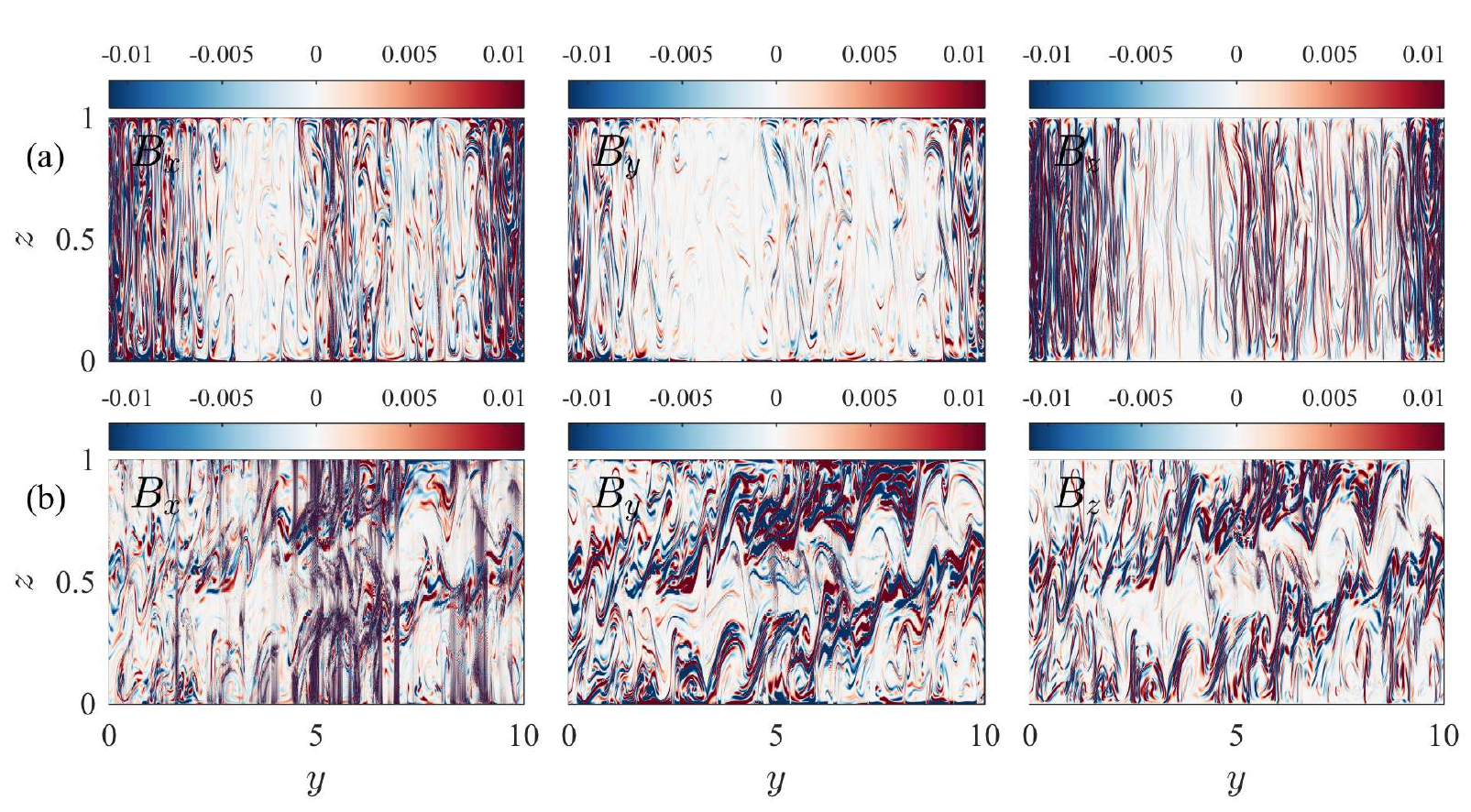}
\caption{\label{fig8} The three magnetic field components for (a) $\phi=\pi/2$, $T_y=0$, $N_{crit}=11.63$, $k_x=5$ and $Rm\sim39000$ and (b) $\phi=\pi/12$, $T_y=-5$, $N_{crit}=19.53$, $k_x=10$ and $Rm\sim35000$. The field has been normalised for each component by its maximum value so that it lies between plus and minus one.}
\end{center}
\end{figure}

Illustrative examples of the form of the magnetic field are given in Figure~\ref{fig8} for cases with and without thermal winds. For  $\phi=\pi/12$, $T_y=-5$, $N_{crit}=19.53$, $k_x=10$ and $Rm\sim35000$ (i.e., a strong thermal wind), Figure~\ref{fig8}(b) shows the spatial form of the magnetic field components at $x=0$. Clearly the field is not uniformly amplified and the regions of strongest field occur, as expected, in regions where the shear is strongest (for reference the snapshots of the flow field for this case are similar to those shown in figure~\ref{fig1}(d))  For comparison, in figure~\ref{fig8}(a), we show the spatial form of the magnetic field components for a case with $\phi=\pi/2$, $T_y=0$, $N_{crit}=11.63$, $k_x=5$ and $Rm\sim39000$ and so there is no thermal wind or tilted rotation (and hence no systematic shear). Here the field small-scale across the layer with no preferred locations for dynamo action.

\begin{figure}
\begin{center}
\includegraphics[trim={1mm 0mm 0mm 0mm}, clip, scale=0.97]{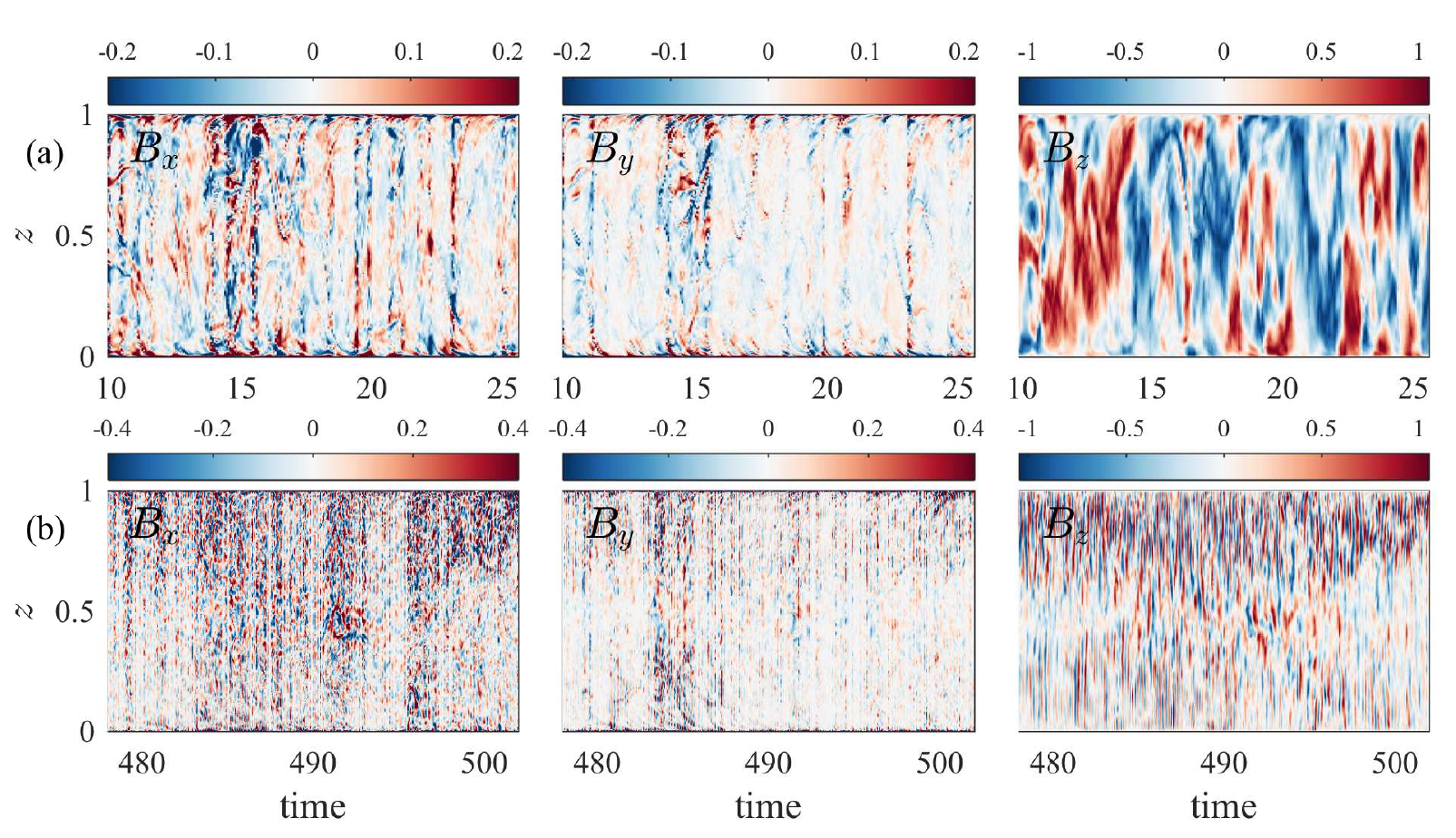}
\caption{\label{fig9} Mean components of the magnetic field at $x=0$ plotted against time and $z$ for the same parameters as used in figure~\ref{fig8}. The exponential growth of the field has been removed at each time. The time has also been rescaled to be measured in units of the turnover time.}
\end{center}
\end{figure}

However, it is of interest to examine the role of the shear in modifying the kinematic large scale properties of the field. To this end we consider the mean ($y$-averaged) fields  as a function of the layer depth and time; recall the field varies sinusoidally in $x$. The plots corresponding to the same parameters as in figure~\ref{fig8} are given in figure~\ref{fig9}. There is little evidence of systematic large scale behaviour, though at any given time the $\langle B_z \rangle$ does appear to have a large-scale component. Hence the dynamo appears to be dominated by small scales as seen in some other high $Rm$ studies.

The plots we include here are representative of the form of the field found for almost all parameter values. Small-scale magnetic fields appear to dominate over the systematic large scales despite the presence of rotation and systematic shear. We shall return to this point in the discussion.


\section{Discussion and Conclusion}\label{conc}
In this paper, we have examined the kinematic dynamo properties of a $2.5$-dimensional Boussinesq convective flow in a rotating Cartesian domain. For these flows a systematic shear is driven naturally either via the interaction of convection with rotation that is not aligned with gravity (as measured by a tilt angle $\phi$) or via the interaction of a latitudinal temperature gradient with rotation (a so-called thermal wind) \citep{hatsom:1983,curthesis:2014}. We have calculated  the hydrodynamic properties ---  including the relative helicity (averaged over half the domain), relative shear and heat transport --- of such flows as a function of tilt angle and Rayleigh number $Ra$. Because of the Boussinesq symmetry, the helicity is an antisymmetric function of height (on average) so the net helicity is zero when averaged over the flow as a whole.

We found that these flows were excellent dynamos (even at high $Rm$), however, even at $Rm\sim30Rm_c$, the growth rate had not reached the asymptotic regime --- this has consequences for our interpretation of dynamo results from three dimensional simulations that purport to explore the high $Rm$ regime. Furthermore, it appears as though these flows act as small-scale dynamos, with very little systematic behaviour being apparent at high $Rm$.

\begin{figure}
\begin{center}
\includegraphics[trim={0mm 0mm 0mm 0mm}, clip, scale=0.95]{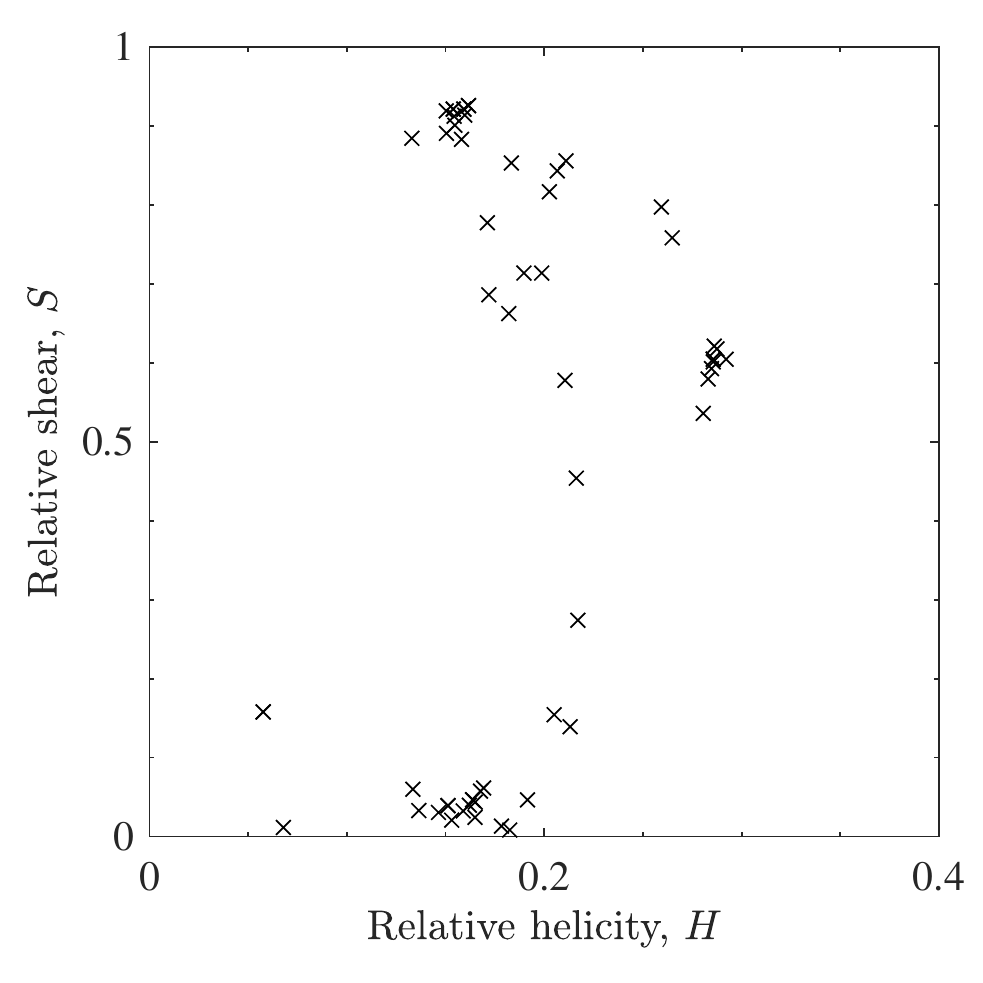}
\caption{\label{fig10} Regime diagram showing the position of all simulations in Relative shear - Relative helicity (averaged over half the depth) space.}
\end{center}
\end{figure}

We conclude by speculating on the reasons for the absence of large-scale dynamo action in our convection system. \citet{pnct:2016} demonstrated that in order for large-scale dynamo waves to be observed at high $Rm$ the underlying flow required to have suitably large (net) relative helicity and shear. Indeed in that paper it was speculated that the presence of large-scale dynamo waves was predicated on the product of the shear and helicity being larger than a critical threshold.
In our system, although the shear is strong, the net helicity is small owing to the Boussinesq symmetry (although the helicity is not insignificant when measured over half the domain -- see figure~\ref{fig10}). It may be that this hinders large-scale dynamo action; indeed in the study of \citet{catthughes:06} the convection had no large scale shear and no net helicity and only small scale dynamo action was found.
It is therefore of interest to examine the dynamo properties of convective systems that allow net helicity to be generated, i.e., those with stratification \citep{curtob:2016}. We propose to extend our investigation to this stratified case in the near future.
We do not believe that breaking the Boussinesq symmetry perturbatively will automatically lead to the detection of a large-scale mode for the magnetic field. Rather it seems from previous results \citep{pnct:2016,npct:2017} that only a flow with a sufficiently large product of shear and net helicity (sometimes characterised by a dynamo number) can yield a large-scale signal that is strong enough to be detected over the small-scale fluctuations. For small helicity and shear there \textit{will} be a large-scale mode, which is detectable by filtering \citep{npct:2017} but this is swamped by the fluctuations. Indeed even for a Boussinesq system with no shear a suitably defined large-scale field can be detected, but this is completely overwhelmed by the fluctuations in the kinematic regime. The presence of shear in combination with a net helicity leads to the formation of propagating waves with a well-defined period --- a signal that is more easily detected among the sea of fluctuations.

We also note that it has been found that the precise nature of the magnetic boundary conditions can make a large difference for the generation of magnetic fields \citep{bushby:2018}, potentially because of the fluxes of helicity that are allowed from the domain \citep{bf:2000,bcmr:2017}. Moreover the problem we have considered here is kinematic,
in reality the Lorentz force acts back on the convective flow and the dynamo properties of the saturated state should be different, potentially saturating the small-scale dynamo and allowing the large-scale dynamo to take over. This can only be examined in a fully three dimensional simulation and work has already started on such a model.

\section*{Acknowledgements}
We thank three anonymous referees for their thoughtful comments that helped to improve the paper. LKC acknowledges support from the European Research Council under ERC grant agreements No. 337705 (CHASM). SMT is supported by STFC grant: ST/N000765/1.
The simulations here were carried out on the University of Exeter supercomputer, a DiRAC Facility jointly funded by STFC, the Large Facilities Capital Fund of BIS and the University of Exeter. This work used the DiRAC Complexity system, operated by the University of Leicester IT Services, which forms part of the STFC DiRAC HPC Facility (www.dirac.ac.uk). This equipment is funded by BIS National E-Infrastructure capital grant ST/K000373/1 and STFC DiRAC Operations grant ST/K0003259/1. DiRAC is part of the National E-Infrastructure.

%%%%%%%%%%%%%%%%%%%%%%%%%%%%%%%%%%%%

\bibliographystyle{gGAF}
%\bibliography{References_all.bib}

\markboth{L.K.~Currie and S.M.~Tobias}{Geoghysical and Astrophysical Fluid Dynamics}
\begin{thebibliography}{38}
\providecommand{\natexlab}[1]{#1}

\bibitem[\protect\citeauthoryear{{Augustson}
  {\itshape{et~al.}}}{2015}]{abmt:2015}
{Augustson}, K., {Brun}, A.S., {Miesch}, M. and {Toomre}, J., {Grand minima and
  equatorward propagation in a cycling stellar convective dynamo}. {\itshape
  Astrophys.~J.} 2015, \textbf{809}, 149.
%(pp.~25).

\bibitem[\protect\citeauthoryear{{Blackman} and {Field}}{2000}]{bf:2000}
{Blackman}, E.G. and {Field}, G.B., {Constraints on the magnitude of {$\alpha$}
  in dynamo theory}. {\itshape Astrophys.~J.} 2000, \textbf{534}, 984--988.

\bibitem[\protect\citeauthoryear{{Bodo} {\itshape{et~al.}}}{2017}]{bcmr:2017}
{Bodo}, G., {Cattaneo}, F., {Mignone}, A. and {Rossi}, P., {Magnetic helicities
  and dynamo action in magneto-rotational turbulence}. {\itshape Astrophys.~J.} 2017,
\textbf{843}, 86.
%(pp.~10).

\bibitem[\protect\citeauthoryear{{Brandenburg} and
  {Subramanian}}{2005}]{bransub:2005}
{Brandenburg}, A. and {Subramanian}, K., {Astrophysical magnetic fields and
  nonlinear dynamo theory}. {\itshape Phys.~Rep.} 2005, \textbf{417},
  1--209.

\bibitem[\protect\citeauthoryear{{Brummell}
  {\itshape{et~al.}}}{1998}]{bht:1998}
{Brummell}, N.H., {Hurlburt}, N.E. and {Toomre}, J., {Turbulent compressible
  convection with rotation. II. Mean flows and differential rotation}.
  {\itshape Astrophys.~J.} 1998, \textbf{493}, 955--969.

\bibitem[\protect\citeauthoryear{{Brun} and
  {Browning}}{2017}]{brunbrowning:2017}
{Brun}, A.S. and {Browning}, M.K., {Magnetism, dynamo action and the
  solar-stellar connection}. {\itshape Living Rev.~Sol.~Phys.} 2017,
\textbf{14}, 4.
%(pp.~133).

\bibitem[\protect\citeauthoryear{{Bushby}
  {\itshape{et~al.}}}{2018}]{bushby:2018}
{Bushby}, P.J., {K{\"a}pyl{\"a}}, P.J., {Masada}, Y., {Brandenburg}, A.,
  {Favier}, B., {Guervilly}, C. and {K{\"a}pyl{\"a}}, M.J., {Large-scale
    dynamos in rapidly rotating plane layer convection}. {\itshape Astron.~Astrophys.} 2018, \textbf{612}, A97.
 % (pp.~16).

\bibitem[\protect\citeauthoryear{{Cattaneo} and {Hughes}}{2006}]{catthughes:06}
{Cattaneo}, F. and {Hughes}, D.W., {Dynamo action in a rotating convective
  layer}. {\itshape  J.~Fluid Mech.} 2006, \textbf{553}, 401--418.

\bibitem[\protect\citeauthoryear{Childress and Gilbert}{1995}]{chilgil:1995}
Childress, S. and Gilbert, A.D., {\itshape Stretch, twist, fold: the fast
  dynamo},  1995 (Berlin: Springer).

\bibitem[\protect\citeauthoryear{{Cowling}}{1933}]{cowling:1933}
{Cowling}, T.G., {The magnetic field of sunspots}. {\itshape Mon.~Not.~R.~Astron.~Soc.} 1933, \textbf{94}, 39--48.

\bibitem[\protect\citeauthoryear{Currie}{2014}]{curthesis:2014}
Currie, L.K., The driving of mean flows by convection. Ph.D. Thesis, University
  of Leeds, 2014.

\bibitem[\protect\citeauthoryear{{Currie} and {Tobias}}{2016}]{curtob:2016}
{Currie}, L.K. and {Tobias}, S.M., {Mean flow generation in rotating anelastic
  two-dimensional convection}. {\itshape Phys.~Fluids} 2016, \textbf{28},
  017101.

\bibitem[\protect\citeauthoryear{Dormy}{2007}]{dormysoward:2007}
Dormy, E. \&~Soward, A.M. (Ed.) {\itshape {Mathematical aspects of natural
  dynamos}},  2007 (CRC Press/Taylor \& Francis).

\bibitem[\protect\citeauthoryear{Galloway and Proctor}{1992}]{Galloway:1992}
Galloway, D.J. and Proctor, M.R.E., Numerical calculations of fast dynamos in
  smooth velocity fields with realistic diffusion. {\itshape Nature} 1992,
  \textbf{356}, 691--693.

\bibitem[\protect\citeauthoryear{{Hathaway} and
  {Somerville}}{1983}]{hatsom:1983}
{Hathaway}, D.H. and {Somerville}, R.C.J., {Three-dimensional simulations of
  convection in layers with tilted rotation vectors}. {\itshape  J.~Fluid Mech.} 1983, \textbf{126}, 75--89.

\bibitem[\protect\citeauthoryear{{Hathaway}
  {\itshape{et~al.}}}{1980}]{htg:1980}
{Hathaway}, D.H., {Toomre}, J. and {Gilman}, P.A., {Convective instability when
  the temperature gradient and rotation vector are oblique to gravity. II -
  Real fluids with effects of diffusion}. {\itshape Geophys.~ Astrophys.~Fluid~Dynam.} 1980, \textbf{15}, 7--37.

\bibitem[\protect\citeauthoryear{{Julien} and
  {Knobloch}}{1998}]{julienknobloch:1998}
{Julien}, K. and {Knobloch}, E., {Strongly nonlinear convection cells in a
  rapidly rotating fluid layer: the tilted f-plane}. {\itshape J.~Fluid Mech.}, 1998, \textbf{360}, 141--178.

\bibitem[\protect\citeauthoryear{{K{\"a}pyl{\"a}} and
  {Brandenburg}}{2009}]{Kapya09}
{K{\"a}pyl{\"a}}, P.J. and {Brandenburg}, A., {Turbulent dynamos with shear and
  fractional helicity}. {\itshape Astrophys.~J.} 2009, \textbf{699}, 1059--1066.

\bibitem[\protect\citeauthoryear{Klapper and Young}{1995}]{Klapper:1995}
Klapper, I. and Young, L.S., Rigorous bounds on the fast dynamo growth-rate
  involving topological entropy. {\itshape Commun.~Math.~Phys.} 1995,
  \textbf{175}, 623--646.

\bibitem[\protect\citeauthoryear{{Krause} and {Raedler}}{1980}]{kr:1980}
{Krause}, F. and {Raedler}, K.H., {\itshape {Mean-field magnetohydrodynamics
  and dynamo theory}},  1980 (Oxford, Pergamon Press).

\bibitem[\protect\citeauthoryear{{Moffatt}}{1978}]{Moffatt:1978}
{Moffatt}, H.K., {\itshape {Magnetic field generation in electrically
  conducting fluids}},  1978 (Cambridge, England, Cambridge
  University Press, 1978 (pp.~353).

\bibitem[\protect\citeauthoryear{{Nigro} {\itshape{et~al.}}}{2017}]{npct:2017}
{Nigro}, G., {Pongkitiwanichakul}, P., {Cattaneo}, F. and {Tobias}, S.M., {What
  is a large-scale dynamo?} {\itshape  Mon.~Not.~R.~Astron.~Soc.} 2017, \textbf{464}, L119--L123.

\bibitem[\protect\citeauthoryear{{Parker}}{1955}]{Parker:1955}
{Parker}, E.N., {Hydromagnetic dynamo models}. {\itshape Astrophys.~J.} 1955,
  \textbf{122}, 293--314.

\bibitem[\protect\citeauthoryear{{Parker}}{1979}]{Parker:1979}
{Parker}, E.N., {\itshape {Cosmical magnetic fields: their origin and their
  activity}},  1979 (Oxford, Clarendon Press; New York, Oxford University
  Press, 1979 (pp.~858).

\bibitem[\protect\citeauthoryear{{Passos} and {Charbonneau}}{2014}]{pc:2014}
{Passos}, D. and {Charbonneau}, P., {Characteristics of magnetic solar-like
  cycles in a 3D MHD simulation of solar convection}. {\itshape Astron.~Astrophys.} 2014,
  \textbf{568}, A113 (pp.~16).

\bibitem[\protect\citeauthoryear{{Pongkitiwanichakul}
  {\itshape{et~al.}}}{2016}]{pnct:2016}
{Pongkitiwanichakul}, P., {Nigro}, G., {Cattaneo}, F. and {Tobias}, S.M.,
{Shear-driven dynamo waves in the fully nonlinear regime}. {\itshape Astrophys.~J.} 2016, \textbf{825}, 23 (pp.~8).

\bibitem[\protect\citeauthoryear{{Ponty}
  {\itshape{et~al.}}}{2001}]{PontGilSow:2001}
{Ponty}, Y., {Gilbert}, A.D. and {Soward}, A.M., {Kinematic dynamo action in
  large magnetic Reynolds number flows driven by shear and convection}.
  {\itshape J. Fluid Mech.} 2001, \textbf{435}, 261--287.

\bibitem[\protect\citeauthoryear{{Roberts}}{1972}]{Roberts:1972}
{Roberts}, G.O., {Dynamo action of fluid motions with two-dimensional
  periodicity}. {\itshape Philos.~T.~Roy.~Soc.~A} 1972, \textbf{271},
  411--454.

\bibitem[\protect\citeauthoryear{{Schekochihin}
  {\itshape{et~al.}}}{2007}]{2007NJPh....9..300S}
{Schekochihin}, A.A., {Iskakov}, A.B., {Cowley}, S.C., {McWilliams}, J.C.,
  {Proctor}, M.R.E. and {Yousef}, T.A., {Fluctuation dynamo and turbulent
  induction at low magnetic Prandtl numbers}. {\itshape New J.~ Phys.} 2007, \textbf{9}, 300--+.

\bibitem[\protect\citeauthoryear{{Sridhar} and {Singh}}{2010}]{Sridhar10}
{Sridhar}, S. and {Singh}, N.K., {The shear dynamo problem for small magnetic
  Reynolds numbers}. {\itshape  J.~Fluid Mech.} 2010, \textbf{664},
  265--285.

\bibitem[\protect\citeauthoryear{{Steenbeck}
  {\itshape{et~al.}}}{1966}]{skr:1966}
{Steenbeck}, M., {Krause}, F. and {R{\"a}dler}, K.-H., {A calculation of the mean electromotive force in an electrically conducting fluid in turbulent motion, under the influence of Coriolis forces}. {\itshape Zeitschrift Naturforschung Teil A} 1966, \textbf{21}, 369--376. [Transl. in pp. 29-47 of: The turbulent dynamo: a translation of a series of papers by F. Krause, K.-H. R{\"a}dler, and M. Steenbeck. Roberts and Stix, 1971. NCAR Technical Note NCAR-TN/60-IA, doi:10.5065/D6DJ5CK7]

\bibitem[\protect\citeauthoryear{{Tobias} and {Cattaneo}}{2008}]{tobcatt:2008a}
{Tobias}, S.M. and {Cattaneo}, F., Dynamo action in complex flows: the quick
  and the fast. {\itshape  J.~Fluid Mech.} 2008, \textbf{601},
  101--122.

\bibitem[\protect\citeauthoryear{{Tobias} and {Cattaneo}}{2013}]{tc:2013}
{Tobias}, S.M. and {Cattaneo}, F., {Shear-driven dynamo waves at high magnetic
  Reynolds number}. {\itshape Nature} 2013, \textbf{497}, 463--465.

\bibitem[\protect\citeauthoryear{{Tobias} and {Cattaneo}}{2015}]{tc:2015}
{Tobias}, S.M. and {Cattaneo}, F., {The electromotive force in multi-scale
  flows at high magnetic Reynolds number}. {\itshape J.~Plasma~Phys.} 2015, \textbf{81}, 395810601.

\bibitem[\protect\citeauthoryear{{Tobias} {\itshape{et~al.}}}{2011}]{tcb:11}
{Tobias}, S.M., {Cattaneo}, F. and {Brummell}, N.H., {On the generation of
  organized magnetic fields}. {\itshape Astrophys.~J.} 2011, \textbf{728}, 153.

\bibitem[\protect\citeauthoryear{Vishik}{1989}]{Vishik:1989}
Vishik, M.M., Magnetic field generation by the motion of a highly conducting
  fluid. {\itshape Geophys.~ Astrophys.~Fluid~Dynam.} 1989, \textbf{48},
  151--167.

\bibitem[\protect\citeauthoryear{{Yousef} {\itshape{et~al.}}}{2008}]{Yousef08}
{Yousef}, T.A., {Heinemann}, T., {Schekochihin}, A.A., {Kleeorin}, N.,
  {Rogachevskii}, I., {Iskakov}, A.B., {Cowley}, S.C. and {McWilliams}, J.C.,
  {Generation of magnetic field by combined action of turbulence and shear}.
  {\itshape Phys.~Rev.~Lett.} 2008, \textbf{100}, 184501.

\end{thebibliography}

\markboth{L.K.~Currie and S.M.~Tobias}{Geophysical and Astrophysical Fluid Dynamics}
\appendices
\section{Heat transport}\label{app1}
\begin{figure}
\begin{center}
\includegraphics[trim={0mm 0mm 0mm 0mm}, clip, scale=1]{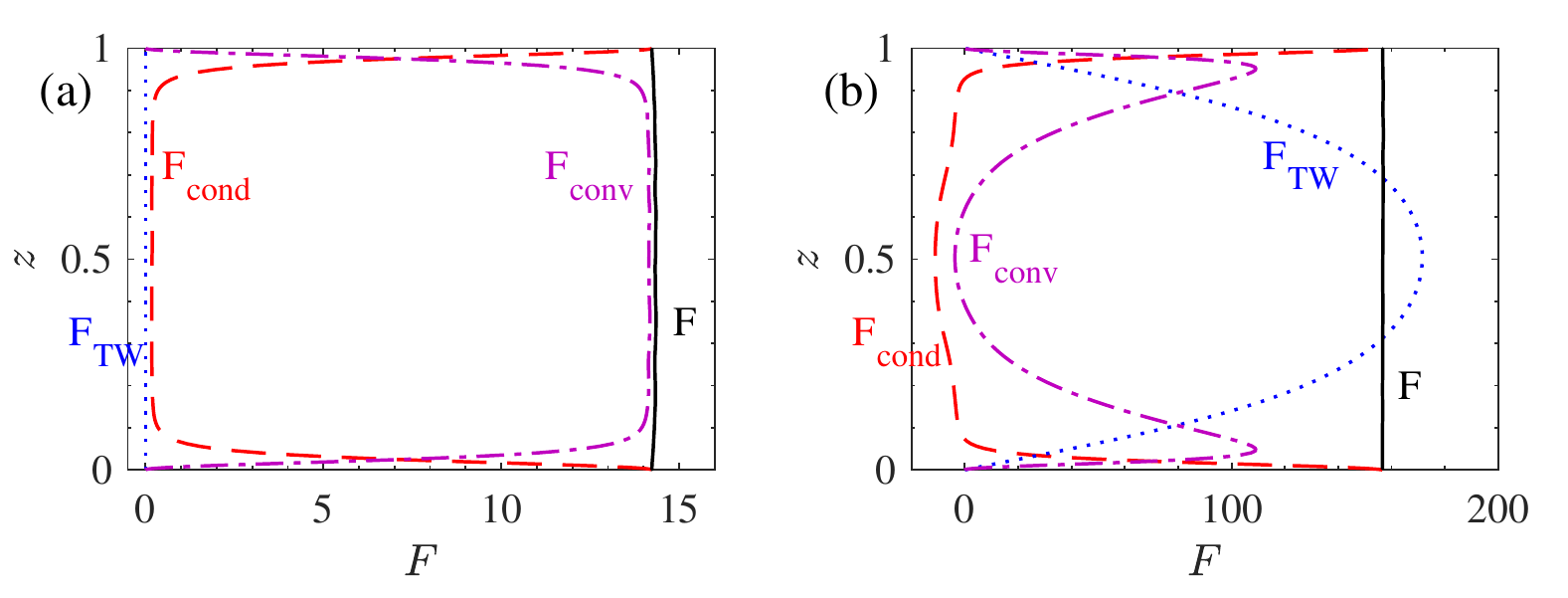}
\caption{\label{figA1}
Components of the heat flux $F_{cond}$ (red, dashed), $F_{conv}$ (purple, dot-dashed), $F_{TW}$ (blue, dotted) and their sum, $F$, (black, solid) for two different cases. Both cases have $Ta=5\times10^6$, $\phi=\pi/12$ and $N_{crit}=11.63$ but in (a) $T_y=0$ and in (b) $T_y=-1$. (Colour online)}
\end{center}
\end{figure}

The heat flux across the layer can be determined by considering the full (non-dimensionalised) heat equation, which can be written as 
\begin{equation}
\frac{\upartial T}{\upartial t} + (\bm{u}{\bm \cdot}\bm\nabla)T\,= \,\nabla^2T\,.
\end{equation}
Then if we assume a statistically-steady state and integrate over the area given by $0\leq y \leq L_y$, $0\leq z\leq z'$ where $0\leq z'\leq 1$ is some depth in the layer we find
\begin{equation}\label{eqvolav}
\int_0^{z'}\int_0^{L_y}\bm\nabla{\bm \cdot}(\bm\nabla T) \, {\mathrm d}y\,{\mathrm d}z\, -\, \int_0^{z'}\int_0^{L_y}\bm\nabla{\bm \cdot}(\bm u T) \, {\mathrm d}y\,{\mathrm d}z\, =\,0\,,
\end{equation}
where we have made use of the incompressibility condition (\ref{incomp}).
Applying the divergence theorem to equation (\ref{eqvolav}) and dividing by $L_y$ to form an equation for the heat flux leads to 
\begin{align}
\frac{1}{L_y}\int_0^{L_y}-\,\frac{\upartial T}{\upartial z}\bigg|_{z=0}\,{\mathrm d}y\, =\, \frac{1}{L_y}\int_0^{L_y}-\,\frac{\upartial T}{\upartial z}\bigg|_{z=z'}\,{\mathrm d}y\, + \,\frac{1}{L_y}\int_0^{L_y} wT\big|_{z=z'}\,{\mathrm d}y\nonumber\\
+\, \frac{1}{L_y}\int_0^{z'} T_yL_yv\big|_{y=L_y}\,{\mathrm d}z\,, \hskip 10mm
\end{align}
which can be written as
\begin{align}
\underbrace{\frac{1}{L_y}\int_0^{L_y}-\,\frac{\upartial T}{\upartial z}\bigg|_{z=0}\,{\mathrm d}y}_{\text{$F_{cond}(z=0)$}}\,= \,\underbrace{\frac{1}{L_y}\int_0^{L_y}-\,\frac{\upartial T}{\upartial z}\bigg|_{z=z'}\,{\mathrm d}y}_{\text{$F_{cond}(z=z')$}}\, + \,\underbrace{\frac{1}{L_y}\int_0^{L_y} w\theta \bigg|_{z=z'}\,{\mathrm d} y}_{\text{$F_{conv}(z=z')$}} \nonumber\\
+ \, \underbrace{\frac{1}{L_y}\int_0^{z'}\!\!\int_0^{L_y}  vT_y\,{\mathrm d}y\, {\mathrm d}z}_{\text{$F_{TW}(z=z')$}} \,.\hskip 10mm
\end{align}
Here $F_{cond}$ defines the heat flux carried by conduction, $F_{conv}$ defines a convective flux and $F_{TW}$ is an additional flux carried by the basic state thermal wind shear; this term is zero when $T_y=0$. There is no internal heat generation in this model and so the flux at the bottom boundary should equal the flux emerging at the top (this is a good diagnostic to check the convergence our simulations). To illustrate how the different components of heat transport vary as a function of depth we have plotted each of these terms for a case with $T_y=0$ (see figure~\ref{figA1}(a)) and a case with similar parameters but $T_y=-1$ (see figure~\ref{figA1}(b)). In (a), we see that (as expected) $F_{TW}=0$, the conductive flux is carrying all the flux near the boundaries but is small in the bulk and $F_{conv}$ is the dominant flux in the interior. The sum of these fluxes is shown by the solid black line, and in a perfectly steady state would be constant across the layer. In (b), $F_{TW}$ now plays a significant role in the heat transport; it carries most of the flux in the mid layers but decreases towards the boundaries. There are small layers where $F_{conv}$ is significant but $F_{cond}$ is only large very close to the boundaries. Unlike in the $T_y=0$ case, $F_{cond}$ actually becomes inward (representing downward transport of heat) at mid depths, that is, the thermal wind has reversed the sign of the temperature gradient in these regions.
The magnitude of the convective flux in the mid region is similar in the two cases shown, however, the total flux $F$ is much larger in the case with a non-zero thermal wind.

\end{document}